\documentclass[aps,prl,twocolumn,groupedaddress,showpacs]{revtex4} 
\usepackage{graphics,epsfig,amsmath,amssymb}
\begin{document}

\title{Maximally entangled mixed states: Creation and concentration}

\author{Nicholas A. Peters} 
\author{Joseph B. Altepeter}
\author{David Branning$^*$\nocite{db}}
\author{Evan R. Jeffrey}
\author{Tzu-Chieh Wei}
\author{Paul G. Kwiat}

\affiliation{Physics Department, University of Illinois, 1110 West Green Street, Urbana, IL 61801}

\date{July 31, 2003}

\begin{abstract}
 Using correlated photons from parametric downconversion, we
    extend the boundaries of experimentally accessible two-qubit Hilbert
    space. 
    Specifically, we have created and characterized maximally
    entangled mixed states (MEMS) that lie above the Werner
    boundary in the linear entropy-tangle plane.  In addition,
    we demonstrate that such states can be efficiently concentrated,
    simultaneously 
increasing both the purity and
    the degree of entanglement.  We investigate a previously unsuspected sensitivity imbalance in common state measures, i.e., the tangle, linear entropy, and fidelity.
\end{abstract}


\maketitle

By exploiting quantum mechanics it is possible to implement provably
secure cryptography~\cite{cryptofocus}, teleportation~\cite{bennett93},
and super-dense coding~\cite{bennett92}.  
These protocols and most others in quantum information processing require
a known initial quantum state, and typically have optimal results for
pure, maximally entangled initial states.  However, decoherence and
dissipation may cause the states to become mixed and/or less entangled.
As the success of a protocol such as quantum teleportation often hinges on
both the purity and the entanglement of the initial state~\cite{bose00},
it is important to study the interplay of these properties.  Using a
source of 2-qubit polarization states~\cite{kwiat99}, we investigate the
creation of maximally entangled mixed states, and their
concentration~\cite{bennett96a,bennett96b,thew01,footnoteNomenclature}.

Entangled states have been demonstrated in a variety of
systems~\cite{rarity90, kwiat95, brendel, sackett00, bowen03, bao03}.
In fact, there are several classes of entangled states; maximally entangled
and nonmaximally entangled pure states~\cite{kwiat95, kwiat99, white99},
nonmaximally entangled mixed states~\cite{white01}, and the special
case of Werner states~\cite{zhang02} (incoherent combinations of
a completely mixed state and a maximally entangled pure state) have all
been experimentally realized using optical qubits.  For some time it was
believed that Werner states possess the most entanglement for a given level
of mixedness.  However, Munro et~al.~\cite{munro01}~discovered a
class of states that are more entangled than Werner states of the 
same purity.  These maximally entangled
mixed states (MEMS) possess the {\it maximal } amount of entanglement
(tangle or entanglement of formation) for a given degree of mixedness
(linear entropy)~\cite{footnotewerner, wei03a}.

By generating states close to the MEMS boundary, we have experimentally explored the region above the Werner state line on the~linear~entropy-tangle plane~\cite{kwiat03}.  We have also implemented a partial-polarizer filtration/concentration technique which simultaneously increases
both purity and entanglement,~at~the~cost~of~decreasing the ensemble size
of initial photon pairs.  Though the implementation requires initial state knowledge, we show that MEMS exist for which this
``Procrustean'' filtering technique~\cite{bennett96a, kwiat01, thew01} is
much more efficient than other recent entanglement
concentration schemes~\cite{pan,tyam03}, even after modification to work on MEMS.

The exact form of the MEMS density matrix depends on the measures used to
quantify the entanglement and mixedness~\cite{wei03a}; here we use the
tangle
($T(\rho)~=~[max\{0,~\lambda_1-\lambda_2-\lambda_3-\lambda_4 \}]^2$)
\cite{wooters98coffman00}, i.e., the concurrence squared;
and the linear entropy
($S_L(\rho)~=~\frac{4}{3}[1-Tr(\rho^2)]$)
\cite{bose00}.
Here $\lambda_i$ are the square roots of the eigenvalues of
$\rho(\sigma_2 \otimes \sigma_2)\rho^{\ast}(\sigma_2 \otimes  \sigma_2)$,
in non-increasing order by magnitude, with 
$\sigma_2= \left( \begin{array}{cc} 0 & -i \\ i & 0 \end{array} \right)$.
For this parameterization, where r is the concurrence, the MEMS density matrices exist in two
subclasses~\cite{munro01}, $\rho_{MEMS~I}$ 
and $\rho_{MEMS~II}$,
 which have two and three eigenvalues, respectively:
%
\begin{eqnarray}
\rho_{MEMS~I}~=~\bordermatrix{ 
& \cr
&\frac{r}{2} &0 &0 &\frac{r}{2} \cr
&0 &1-r &0 &0 \cr
&0 &0 &0 &0 \cr
&\frac{r}{2} &0 &0 &\frac{r}{2} \cr}
 ,~~~~\frac{2}{3} \leq r \leq 1,
\label{memsI} \\
\rho_{MEMS~II}~=~\bordermatrix{ 
& \cr
&\frac{1}{3} &0 &0 &\frac{r}{2} \cr
&0 &\frac{1}{3} &0 &0 \cr
&0 &0 &0 &0 \cr
&\frac{r}{2} &0 &0 &\frac{1}{3} \cr}
 ,~~~~0 \leq r \leq \frac{2}{3}.
\label{memsII}
\end{eqnarray}
\begin{figure}
\begin{center}
\includegraphics[width=8.6cm]{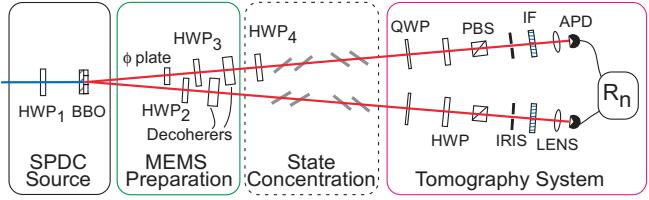}
\vspace{-.5cm}
\tiny  \caption{ Experimental arrangement to create,
and concentrate MEMS.  A half-waveplate
(HWP$_1$) sets the initial entanglement of the pure state.  The $\phi$-plate
sets the relative phase between $|HH\rangle$ and $|VV\rangle$ in the
initial state.  HWP$_2$ and HWP$_3$ rotate the state into the active bases
of the decoherers to adjust the amount of entropy.  The tomography system
uses a quarter-waveplate (QWP), HWP, and a polarizer in each arm to analyze in arbitrary
polarization bases; the transmitted photons are counted in coincidence via
avalanche photodiodes.  The dashed box contains HWP$_4$ (oriented to rotate $|H\rangle \leftrightarrow |V\rangle$ in the first arm of the experiment) and concentrating elements (a
variable number of glass pieces oriented at Brewster's angle to
completely transmit $|H\rangle$, but only partially transmit $|V\rangle$). }
\vspace{-1cm}
\label{memsexp}
\end{center}
\end{figure}

Our creation of MEMS involves three steps: creating an initial
state of arbitrary entanglement, applying local unitary transformations,
and inducing decoherence.  First, frequency degenerate 702-nm photons
are created by pumping thin nonlinear $\beta$-Barium Borate (BBO) crystals with a 351-nm
Ar-ion laser.  Polarization entanglement is realized by pumping two
such crystals oriented such that their optic axes are in perpendicular
planes.  With a pump polarized at $\theta_1$, a variable entanglement
superposition state $\cos\theta_1 |HH\rangle+\sin\theta_1 |VV\rangle$
is created, where $|HH\rangle$ represents two horizontally polarized and
$|VV\rangle$ two vertically polarized photons \cite{kwiat99, white99}.
The pump polarization is controlled by a half-wave plate (HWP$_1$ in
Fig.~\ref{memsexp}) set to $\theta_1/2$.  

To create the MEMS I, we start by setting the initial degree of entanglement to that
of the target MEMS.  Next a maximum likelihood tomography~\cite{white99,
james01} of this initial entangled state is taken and used to numerically determine the appropriate settings of HWP$_2$ and HWP$_3$ in
Fig.~\ref{memsexp}. 
These waveplates set the diagonal elements of the density
matrix to 
the target values for the desired MEMS. The initial tomography must be 
precise, because the waveplate settings are critically dependent on the
initial state, as well as on the precise birefringent retardation of
the waveplates themselves.  After the waveplates, the state passes
through decoherers, which lower specific off-diagonal elements
in the density matrix, yielding the final state.  In our scheme,
each decoherer is a thick birefringent element ($\sim$1~cm quartz, with optic axis horizontal)
chosen to have a polarization-dependent optical path length difference
($\sim$140$\lambda$~\cite{footnotedech}) greater than the downconverted photons'
coherence length ($L_c \equiv \lambda^2/\Delta \lambda\cong 70\lambda$, determined by
a 10-nm FWHM interference filter placed before each detector), but much
less than the coherence length of the pump~\cite{berg}.  

\begin{figure}[!t]
\begin{center}
\includegraphics[width=7.3cm]{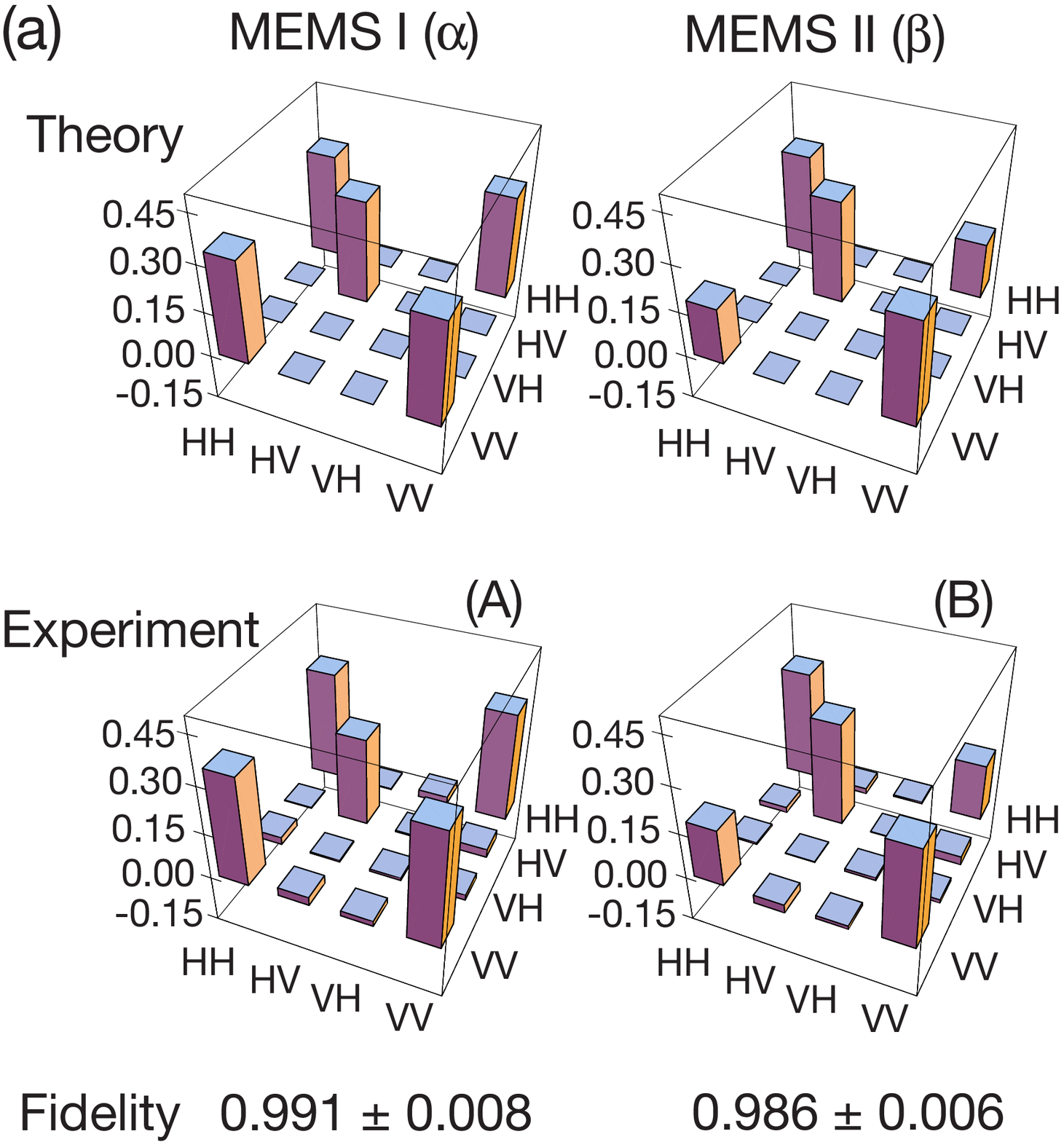}
\includegraphics[width=8cm]{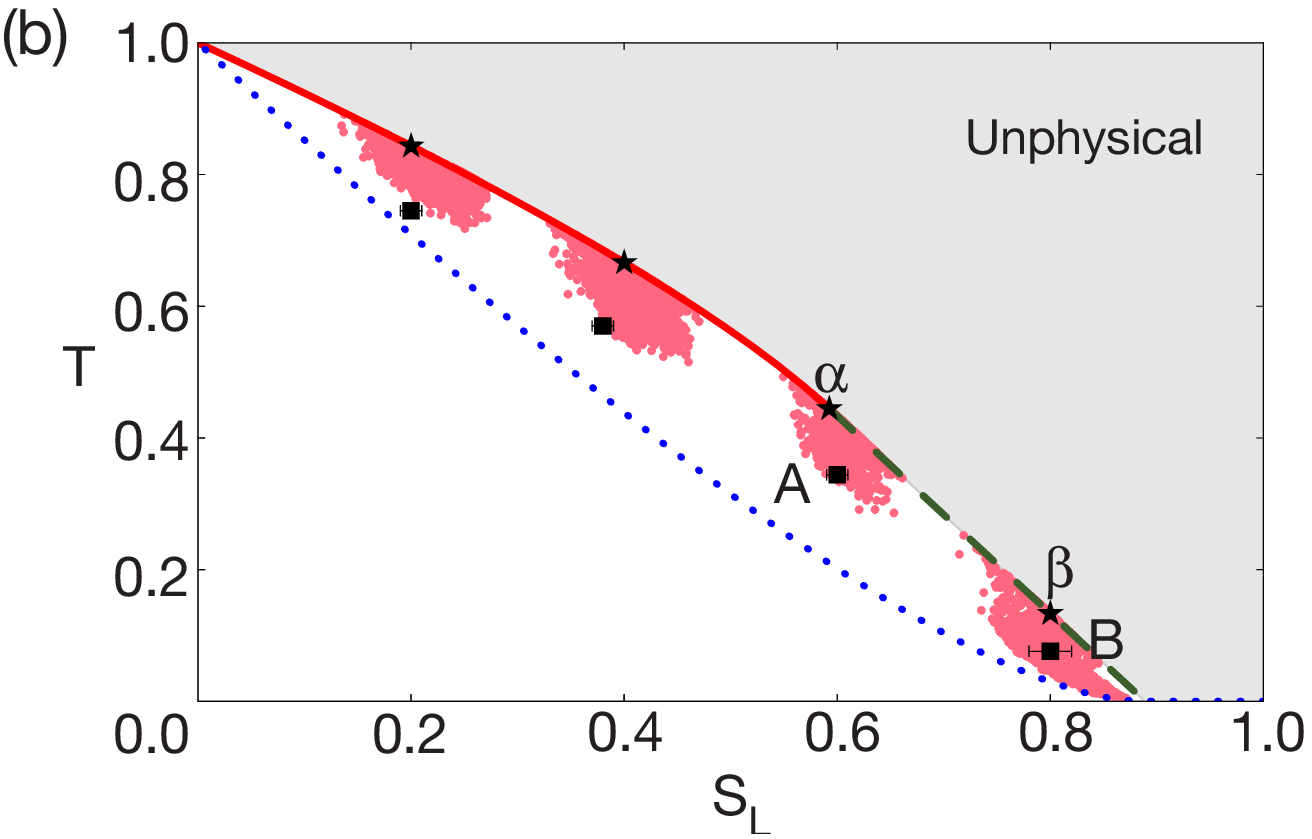}
\vspace{-0.4cm}
\tiny  \caption{MEMS data.  (a) Density matrix plots of the real
components of a MEMS I ($r=\frac{2}{3}$) and a MEMS II ($r=0.3651$).
The imaginary components are negligible (on average less than 0.02) and not
shown.  (b) Linear entropy-tangle plane. Shown are theoretical curves
for MEMS I (solid line), MEMS II (dashed line), and Werner states (dotted
line).  Four target MEMS are indicated by stars; experimental
realizations are shown as squares with error bars.
The shaded patches around each target state show the tangle ($T$)
and linear entropy ($S_L$) for 5000 numerically generated density matrices
that have at least 0.99 fidelity~\cite{jozsa94} with the target state.  $T$=0 (1) corresponds to a product (maximally entangled) state.  $S_L$=0 (1) corresponds to a pure (completely mixed) state.}
\label{memsdata}
\vspace{-1cm}
\end{center}
\end{figure}

The decoherer
in each arm couples the polarization with the relative arrival times
of the photons~\cite{barbieri03}.  While two horizontal ($|HH\rangle$) or two vertical
($|VV\rangle$) photons will be detected at the same time,
the state $|HV\rangle$ will in principle 
be detected first in arm one and then in arm
two, and vice versa for the state $|VH\rangle$ (assuming the decoherer
slows vertically polarized photons relative to horizontally polarized
ones).  Tracing over timing information during state analysis then erases
coherence between any distinguishable terms of the state (i.e., only the coherence term between $|HH\rangle$ and $|VV\rangle$ remains).
A sample tomography of a MEMS I is shown in Fig.~\ref{memsdata}(a).

MEMS II are created by first producing the MEMS I at the
MEMS I/II boundary, i.e., the state with $r=\frac{2}{3}$.  In order
to travel along the MEMS II curve, the optical path length difference
in {\it one} arm must be varied from 140$\lambda$. This couples different
relative timings to the $|HH\rangle$ and $|VV\rangle$ states, reducing
the coherence between them.  For instance, to make the MEMS II (B)
in Fig.~\ref{memsdata}(a), 140$\lambda$ decoherence was used in one
arm, 90$\lambda$ in the other.  Fig.~\ref{memsdata}(a) indicates very good agreement between theory and experiment with fidelities of $\sim$99\% (the fidelity~\cite{jozsa94} between the target state $\rho_t$ and
the measured state $\rho_m$ is given by $F(\rho_t,\rho_m)\equiv \left| {\rm
Tr}\left(\sqrt{\sqrt{\rho_t}\rho_m\sqrt{\rho_t}}\right) \right|^2
\label{fidelity}$).

The states (A) and (B) are shown in the $S_L$-$T$ plane in
Fig.~\ref{memsdata}(b), along with other MEMS we created.  The states do
not hit their $S_L$-$T$ targets (shown as stars in the figure) within
errors, even though the states have very high fidelities ($ \gtrsim
99\%$) with their respective targets.  To explore the discrepancy, for
each target we numerically generated 5000 density matrices that had at
least 0.99 fidelity with the target density matrix.  The $S_L$ and $T$
of the numerically generated states are plotted in Fig.~\ref{memsdata}(b)
as shaded regions surrounding the  targets.  The fact that these regions
are rather large (and overlap with our measured MEMS) explains our results,
but is surprising nonetheless.  The unexpectedly large size of these
patches arises from the great difference in sensitivity between the state
measures of fidelity, tangle and entropy: for small perturbations ($\delta r$) of the MEMS parameter $r$, the fidelity is only quadratic in $\delta r$, while $S_L$ and $T$ are linear in $\delta r$~\cite{footnotesensitivity}.

\begin{figure}
\begin{center}
\includegraphics[width=8.6cm]{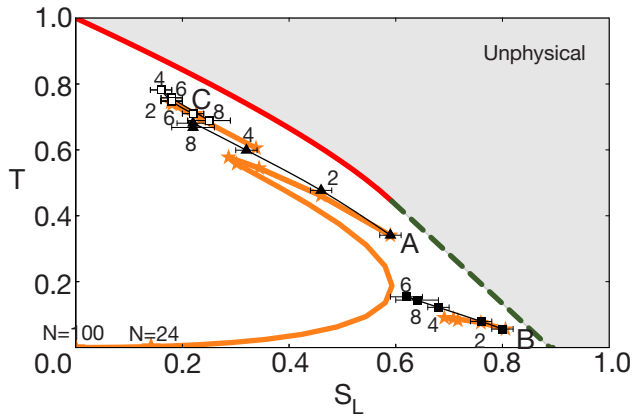} 
\vspace{-.6cm}
\tiny  \caption{Concentration data.  Shown are concentrations for three
initial states, A (triangles) and B (filled squares) as in Fig.~\ref{memsdata}, and C (open squares), along with the number of partial polarizing glass
pieces in each arm.  The expected concentrated state path, calculated
using~\cite{thew01}, is shown with stars.  The concentrated
states agree with theory for small numbers of glass pieces, but as
more slips are used, the state concentrates better than expected.  We believe this is due to the extreme sensitivity of the trajectory to small changes in the initial state.  However, even in theory,
excessive filtration will eventually produce a pure product state
(shown as an extension of A's theory curve), due to small errors in the
initial MEMS.}
\label{concentrationdata}
\vspace{-1cm}
\end{center}
\end{figure}

While our initial goal was to produce states of maximal tangle for a given
linear entropy, maximally entangled pure states are
generally more useful for quantum information protocols.  However, in some
cases, weakly entangled mixed states may be the only available resource.
It is therefore important to study ways to simultaneously decrease the
entropy and increase the entanglement of an ensemble of photon pairs
(necessarily at the cost of reducing the size of the ensemble).  Recently several
such entanglement concentration experiments have been reported, relying
on two-photon interference effects~\cite{tyam03, pan}.  An interesting
characteristic of MEMS is that they can be readily concentrated by a
``Procrustean'' method of local filtering \cite{bennett96a, kwiat01}.
To concentrate we first modify the MEMS using HWP$_4$ at $45^\circ$ to exchange $|H\rangle \leftrightarrow |V\rangle$ in the first arm, changing the non-zero diagonal elements of the MEMS density matrix to $|HV\rangle\langle HV|$, $|VH\rangle\langle VH|$, and $|VV\rangle\langle VV|$.  By reducing the $|VV\rangle \langle VV|$ element of the rotated MEMS, the outcome will be driven
toward the maximally entangled pure state $|\phi^{+}\rangle\equiv
(|HV\rangle+|VH\rangle)/\sqrt{2}$.  We achieve this by inserting glass
pieces (each piece consisting of four $\sim$1mm thick microscope slides sandwiched together with index matching fluid) oriented at Brewster's angle, as indicated in the
dotted box in Fig.~\ref{memsexp}.  Equal numbers of pieces are used in both arms; they are oriented to nearly perfectly transmit horizontally polarized photons (transmission probability $T_{H}=0.990\pm 0.006$)
while partially reflecting vertically polarized photons ($T_{V}=0.740\pm
0.002$).

We concentrated a variety of MEMS.  Fig.~\ref{concentrationdata} shows
the results for the MEMS I and II of Fig.~\ref{memsdata} and
an additional MEMS I (C).  As the number of glass pieces is increased,
the states initially become more like a pure maximally entangled
state.  For example, in the case of (A), the
fidelity of the initial MEMS with the state $|\phi^{+}\rangle$ is 0.672.
When the state is concentrated with eight glass slips per arm, the
fidelity with $|\phi^{+}\rangle$ is 0.902; 4.5\%
of the initial photon pairs survive this filtering process.  The theoretical maximum survival probability is 6.4\%.  Note a characteristic difference between the two MEMS subclasses: MEMS II cannot be filtered into a Bell state.

We now compare the theoretical efficiency of our local filtering
scheme with the interference-based concentration proposal of Bennett et al.~\cite{bennett96b}, assuming identical initial MEMS and
the same number of photon pairs.  We shall compare the average
 final entanglement of formation ($E_F$)\cite{wooters98coffman00} (i.e., the $E_F$ of the concentrated
 state multiplied by the probability of
success) {\it per initial pair\/}.  The Bennett el al.~\cite{bennett96b} scheme was recently approximated by Pan et al.~\cite{pan}, with CNOT operations replaced by polarizing beam splitters; however, due to incomplete Bell state
analysis, the probability of successful concentration is only
50\% of the original proposal (the recent scheme of Yamamoto et al. \cite{tyam03} 
is unable to distill MEMS).  The first step of both schemes
is to perform a ``twirling'' operation~\cite{twirl} to transform a general entangled
state into a Werner state.  However, this initial operation usually {\it decreases} the 
entanglement, and the scheme with twirling is
efficient only when $r$ is close to 1.

In fact, MEMS I could also be distilled {\it without} the twirling operation, using the scheme of Pan et al.  But then the probability of success depends on
the parameter $r$.  
\begin{table}[t]
\begin{center}
\begin{tabular}{||c|c|c|c|c||}
\hline  
\begin{tabular}{c}
{\rm Concent.}\\
{\rm method} 
\end{tabular}
   & \begin{tabular}{c}
   {\rm Prob. of}\\
   {\rm  success}
   \end{tabular} 
   &\begin{tabular}{c}
   {\rm $E_F$ when}\\
   {\rm  successful}
   \end{tabular} 
    & \begin{tabular}{c}
    {\rm Ideal $E_F$}\\
    {\rm per pair}
    \end{tabular}  & \begin{tabular}{c}
    {\rm Exp. $E_F$}\\
    {\rm per pair}
    \end{tabular} \\\hline
\begin{tabular}{c}
   {\rm Twirling~\cite{bennett96b}}
   \end{tabular}    & 74.8\% & 0.51 & 0.19 & NA \\ \hline 
\begin{tabular}{c}
   {\rm No Twirling~\cite{pan}}
   \end{tabular}  & 35.2\% & 0.80 & 0.14 & $\lesssim 10^{-5}$ \\ \hline
 {\rm Procrustean} & \phantom{s} & \phantom{s} & \phantom{s}& \phantom{s}
 \\
 {\rm 2 pieces} & 50.4\% & 0.81 & 0.41& 0.14 \\
 {\rm 4 pieces}  & 26.4\% & 0.88 & 0.23 &0.07\\
 {\rm 6 pieces}  & 14.2\% & 0.93 & 0.13 &0.03\\
 
\hline
\end{tabular}
\end{center}
\vspace{-.5cm}
\caption{\label{tb:comparison1} Efficiency comparison of concentration technique of Bennett et al. using ideal CNOT~\cite{bennett96b},
interference-based concentration~\cite{pan} without twirling, and Procrustean filtering,
for an initial MEMS with $r=0.778$ and $E_F=0.69$.
The scheme of Bennett et al. requires a twirling operation that decreases the initial $E_F$ to 0.418 before the concentration~\cite{twirl}.  
In all schemes, except for the final column, we assume the ideal case, i.e., no loss and perfect
detector
efficiency. To calculate the no-loss result for our filtering scheme, we normalize the measured partial polarizer transmission coefficients (of a single glass piece) to $T_H=0.740/0.990$ and $T_V=1$. 
In the interference schemes, columns 2-4 assume the existence of the required two identical pairs, but in practice this requirement is difficult to achieve~\cite{footnoteeff}.  This limitation is reflected in column 5, which lists the average $E_F$ per initial pair achieved in our experiment, to be compared with the much lower value achievable with current interference method technology.}
\vspace{-.5cm}
\end{table}
For most MEMS, the maximum distillation efficiency from
filtration can exceed that achievable using the interference-based
methods~\cite{footnoteeff}.  For example, as shown in Table I, when the
initial state is a MEMS with $r=0.778$, the two-piece filtering technique has
a theoretical $E_F$ per pair nearly three times higher than the interference scheme without twirling, even though a successful concentration produces nearly the same $E_F$.  In theory, using 2 to 5 slips achieves both higher
entanglement of the successful state and better average entanglement
yield.  In practice, the filtration technique is {\it much} more efficient (see the final columns of Table I)~\cite{footnoteeff}.

We have demonstrated a tunable source of high fidelity MEMS.
As a consequence of comparing the $T$-$S_L$ and fidelity values of
generated MEMS with the theoretical targets, we identify and explain an unsuspected
difference in sensitivity in these state measures.  Furthermore, we have applied a Procrustean filtering technique to several MEMS, realizing a
measured efficiency that is well above that achievable using other methods.
However, in the limit of very strong filtering, small perturbations in
the initial state will eventually dominate the process, yielding only
product states (see Fig. 3).  In practice, therefore, it may be optimal
to combine both methods.

This work was supported by the DCI Postdoctoral Fellowship
Program, ARDA and NSF grant $\#$ EIA-0121568.
\vspace{-.7cm}
\bibliography{mems}

\begin{thebibliography}{34}
\expandafter\ifx\csname natexlab\endcsname\relax\def\natexlab#1{#1}\fi
\expandafter\ifx\csname bibnamefont\endcsname\relax
  \def\bibnamefont#1{#1}\fi
\expandafter\ifx\csname bibfnamefont\endcsname\relax
  \def\bibfnamefont#1{#1}\fi
\expandafter\ifx\csname citenamefont\endcsname\relax
  \def\citenamefont#1{#1}\fi
\expandafter\ifx\csname url\endcsname\relax
  \def\url#1{\texttt{#1}}\fi
\expandafter\ifx\csname urlprefix\endcsname\relax\def\urlprefix{URL }\fi
\providecommand{\bibinfo}[2]{#2}
\providecommand{\eprint}[2][]{\url{#2}}

\bibitem[*]{db}Present address:$\,${\it Dept. of Physics and Optical Eng., Rose-Hulman~Inst.~of~Tech.,~Terre~Haute,~Indiana,~47803. }

\bibitem[{cry(2002)}]{cryptofocus}
\emph{\bibinfo{title}{{\rm N.~Gisin, G.~Ribordy, W.~Tittel, and H.~Zbinden,
  Rev. Mod. Phys. {\bf 74}, 145 (2002); Focus on Quantum Cryptography, New J.
  Phys. {\bf 4}, 41 }}} (\bibinfo{year}{2002}).

\bibitem[{\citenamefont{Bennett{ \it et~al.}}(1993)}]{bennett93}
\bibinfo{author}{\bibfnamefont{C.~H.} \bibnamefont{Bennett{ \it et~al.}}},
  \bibinfo{journal}{Phys.~Rev.~Lett.} \textbf{\bibinfo{volume}{70}},
  \bibinfo{pages}{1895} (\bibinfo{year}{1993}).

\bibitem[{\citenamefont{Bennett and Wiesner}(1992)}]{bennett92}
\bibinfo{author}{\bibfnamefont{C.~H.} \bibnamefont{Bennett}} \bibnamefont{and}
  \bibinfo{author}{\bibfnamefont{S.~J.} \bibnamefont{Wiesner}},
  \bibinfo{journal}{Phys.~Rev.~Lett.} \textbf{\bibinfo{volume}{69}},
  \bibinfo{pages}{2881} (\bibinfo{year}{1992}).

\bibitem[{\citenamefont{Bose and Vedral}(2000)}]{bose00}
\bibinfo{author}{\bibfnamefont{S.}~\bibnamefont{Bose}} \bibnamefont{and}
  \bibinfo{author}{\bibfnamefont{V.}~\bibnamefont{Vedral}},
  \bibinfo{journal}{Phys.~Rev.~A} \textbf{\bibinfo{volume}{61}},
  \bibinfo{pages}{R040101} (\bibinfo{year}{2000}).

\bibitem[{\citenamefont{Kwiat{ \it et al.}}(1999)}]{kwiat99}
\bibinfo{author}{\bibfnamefont{P.~G.} \bibnamefont{Kwiat{ \it et al.}}},
  \bibinfo{journal}{Phy.~Rev.~A} \textbf{\bibinfo{volume}{60}},
  \bibinfo{pages}{R773} (\bibinfo{year}{1999}).

\bibitem[{\citenamefont{Bennett et~al.}(1996)\citenamefont{Bennett, Bernstein,
  Popescu, and Schumacher}}]{bennett96a}
\bibinfo{author}{\bibfnamefont{C.~H.} \bibnamefont{Bennett}},
  \bibinfo{author}{\bibfnamefont{H.~J.} \bibnamefont{Bernstein}},
  \bibinfo{author}{\bibfnamefont{S.}~\bibnamefont{Popescu}}, \bibnamefont{and}
  \bibinfo{author}{\bibfnamefont{B.}~\bibnamefont{Schumacher}},
  \bibinfo{journal}{Phys.~Rev.~A} \textbf{\bibinfo{volume}{53}},
  \bibinfo{pages}{2046} (\bibinfo{year}{1996}).

\bibitem[{\citenamefont{Bennett{ \it et~al.}}(1996)}]{bennett96b}
\bibinfo{author}{\bibfnamefont{C.~H.} \bibnamefont{Bennett{ \it et~al.}}},
  \bibinfo{journal}{Phys.~Rev.~Lett.} \textbf{\bibinfo{volume}{76}},
  \bibinfo{pages}{722} (\bibinfo{year}{1996}).

\bibitem[{\citenamefont{Thew and Munro}(2001)}]{thew01}
\bibinfo{author}{\bibfnamefont{R.~T.} \bibnamefont{Thew}} \bibnamefont{and}
  \bibinfo{author}{\bibfnamefont{W.~J.} \bibnamefont{Munro}},
  \bibinfo{journal}{Phys.~Rev.~A} \textbf{\bibinfo{volume}{63}},
  \bibinfo{pages}{R030302} (\bibinfo{year}{2001}).

\bibitem[{foo({\natexlab{a}})}]{footnoteNomenclature}
\emph{\bibinfo{title}{{\rm After~\cite{thew01}, we use ``concentration'' to
  indicate an increase of both purity and entanglement.}}}

\bibitem[{\citenamefont{Rarity and Tapster}(1990)}]{rarity90}
\bibinfo{author}{\bibfnamefont{J.~G.} \bibnamefont{Rarity}} \bibnamefont{and}
  \bibinfo{author}{\bibfnamefont{P.~R.} \bibnamefont{Tapster}},
  \bibinfo{journal}{Phys.~Rev.~Lett.} \textbf{\bibinfo{volume}{64}},
  \bibinfo{pages}{2495} (\bibinfo{year}{1990}).

\bibitem[{\citenamefont{Kwiat{ \it et al.}}(1995)}]{kwiat95}
\bibinfo{author}{\bibfnamefont{P.~G.} \bibnamefont{Kwiat{ \it et al.}}},
  \bibinfo{journal}{Phys.~Rev.~Lett.} \textbf{\bibinfo{volume}{75}},
  \bibinfo{pages}{4337} (\bibinfo{year}{1995}).

\bibitem[{\citenamefont{Brendel et~al.}(1999)\citenamefont{Brendel, Gisin,
  Tittel, and Zbinden}}]{brendel}
\bibinfo{author}{\bibfnamefont{J.}~\bibnamefont{Brendel}},
  \bibinfo{author}{\bibfnamefont{N.}~\bibnamefont{Gisin}},
  \bibinfo{author}{\bibfnamefont{W.}~\bibnamefont{Tittel}}, \bibnamefont{and}
  \bibinfo{author}{\bibfnamefont{H.}~\bibnamefont{Zbinden}},
  \bibinfo{journal}{Phys.~Rev.~Lett.} \textbf{\bibinfo{volume}{82}},
  \bibinfo{pages}{2594} (\bibinfo{year}{1999}).

\bibitem[{\citenamefont{Sackett{ \it et~al.}}(2000)}]{sackett00}
\bibinfo{author}{\bibfnamefont{C.~A.} \bibnamefont{Sackett{ \it et~al.}}},
  \bibinfo{journal}{Nature} \textbf{\bibinfo{volume}{404}},
  \bibinfo{pages}{256} (\bibinfo{year}{2000}).

\bibitem[{\citenamefont{Bowen et~al.}(2003)\citenamefont{Bowen, Schnabel, Lam,
  and Ralph}}]{bowen03}
\bibinfo{author}{\bibfnamefont{W.~P.} \bibnamefont{Bowen}},
  \bibinfo{author}{\bibfnamefont{R.}~\bibnamefont{Schnabel}},
  \bibinfo{author}{\bibfnamefont{P.~K.} \bibnamefont{Lam}}, \bibnamefont{and}
  \bibinfo{author}{\bibfnamefont{T.~C.} \bibnamefont{Ralph}},
  \bibinfo{journal}{Phys.~Rev.~Lett.} \textbf{\bibinfo{volume}{90}},
  \bibinfo{pages}{043601} (\bibinfo{year}{2003}).

\bibitem[{\citenamefont{Bao et~al.}(2003)\citenamefont{Bao, Bragas, Furdyna,
  and Merlin}}]{bao03}
\bibinfo{author}{\bibfnamefont{J.}~\bibnamefont{Bao}},
  \bibinfo{author}{\bibfnamefont{A.~V.} \bibnamefont{Bragas}},
  \bibinfo{author}{\bibfnamefont{J.~K.} \bibnamefont{Furdyna}},
  \bibnamefont{and} \bibinfo{author}{\bibfnamefont{R.}~\bibnamefont{Merlin}},
  \bibinfo{journal}{Nature Materials} \textbf{\bibinfo{volume}{2}},
  \bibinfo{pages}{175} (\bibinfo{year}{2003}).

\bibitem[{\citenamefont{White et~al.}(1999)\citenamefont{White, James,
  Eberhard, and Kwiat}}]{white99}
\bibinfo{author}{\bibfnamefont{A.~G.} \bibnamefont{White}},
  \bibinfo{author}{\bibfnamefont{D.~F.~V.} \bibnamefont{James}},
  \bibinfo{author}{\bibfnamefont{P.~H.} \bibnamefont{Eberhard}},
  \bibnamefont{and} \bibinfo{author}{\bibfnamefont{P.~G.} \bibnamefont{Kwiat}},
  \bibinfo{journal}{Phys.~Rev.~Lett.} \textbf{\bibinfo{volume}{83}},
  \bibinfo{pages}{3103} (\bibinfo{year}{1999}).

\bibitem[{\citenamefont{White et~al.}(2001)\citenamefont{White, James, Munro,
  and Kwiat}}]{white01}
\bibinfo{author}{\bibfnamefont{A.~G.} \bibnamefont{White}},
  \bibinfo{author}{\bibfnamefont{D.~F.~V.} \bibnamefont{James}},
  \bibinfo{author}{\bibfnamefont{W.~J.} \bibnamefont{Munro}}, \bibnamefont{and}
  \bibinfo{author}{\bibfnamefont{P.~G.} \bibnamefont{Kwiat}},
  \bibinfo{journal}{Phys.~Rev.~A} \textbf{\bibinfo{volume}{65}},
  \bibinfo{pages}{012301} (\bibinfo{year}{2001}).

\bibitem[{\citenamefont{Zhang et~al.}(2002)\citenamefont{Zhang, Huang, Li, and
  Guo}}]{zhang02}
\bibinfo{author}{\bibfnamefont{Y.~S.} \bibnamefont{Zhang}},
  \bibinfo{author}{\bibfnamefont{Y.~F.} \bibnamefont{Huang}},
  \bibinfo{author}{\bibfnamefont{C.~F.} \bibnamefont{Li}}, \bibnamefont{and}
  \bibinfo{author}{\bibfnamefont{G.~C.} \bibnamefont{Guo}},
  \bibinfo{journal}{Phys.~Rev.~A} \textbf{\bibinfo{volume}{66}},
  \bibinfo{pages}{062315} (\bibinfo{year}{2002}).

\bibitem[{\citenamefont{Munro et~al.}(2001)\citenamefont{Munro, James, White,
  and Kwiat}}]{munro01}
\bibinfo{author}{\bibfnamefont{W.~J.} \bibnamefont{Munro}},
  \bibinfo{author}{\bibfnamefont{D.~F.~V.} \bibnamefont{James}},
  \bibinfo{author}{\bibfnamefont{A.~G.} \bibnamefont{White}}, \bibnamefont{and}
  \bibinfo{author}{\bibfnamefont{P.~G.} \bibnamefont{Kwiat}},
  \bibinfo{journal}{Phys.~Rev.~A} \textbf{\bibinfo{volume}{64}},
  \bibinfo{pages}{R030302} (\bibinfo{year}{2001}).

\bibitem[{foo({\natexlab{b}})}]{footnotewerner}
\emph{\bibinfo{title}{{\rm Note that for certain entanglement and mixedness
  parameterizations, the Werner states {\it are} the MEMS~\cite{wei03a}}}}.

\bibitem[{\citenamefont{Wei{ \it et~al.}}(2003)}]{wei03a}
\bibinfo{author}{\bibfnamefont{T.~C.} \bibnamefont{Wei{ \it et~al.}}},
  \bibinfo{journal}{Phys.~Rev.~A} \textbf{\bibinfo{volume}{67}},
  \bibinfo{pages}{022110} (\bibinfo{year}{2003}).

\bibitem[{kwi(2003)}]{kwiat03}
\emph{\bibinfo{title}{{\rm P.~Kwiat{ \it et~al.}, quant-ph/0303040}}}
  (\bibinfo{year}{2003}).

\bibitem[{\citenamefont{Kwiat et~al.}(2001)\citenamefont{Kwiat, Barraza-Lopez,
  Stefanov, and Gisin}}]{kwiat01}
\bibinfo{author}{\bibfnamefont{P.~G.} \bibnamefont{Kwiat}},
  \bibinfo{author}{\bibfnamefont{S.}~\bibnamefont{Barraza-Lopez}},
  \bibinfo{author}{\bibfnamefont{A.}~\bibnamefont{Stefanov}}, \bibnamefont{and}
  \bibinfo{author}{\bibfnamefont{N.}~\bibnamefont{Gisin}},
  \bibinfo{journal}{Nature} \textbf{\bibinfo{volume}{409}},
  \bibinfo{pages}{1014} (\bibinfo{year}{2001}).

\bibitem[{pan()}]{pan}
\emph{\bibinfo{title}{{\rm J.~W.~Pan, C.~Simon, \v{C}.~Brukner, and
  A.~Zeilinger, Nature {\bf 410}, 1067 (2001); J.~W.~Pan {\it et al.}, Nature
  {\bf 423} 417 (2003); Z.~Zhao {\it et al.}, Phys.~Rev.~Lett. {\bf 90} 207901
  (2003).}}}

\bibitem[{\citenamefont{Yamamoto et~al.}(2003)\citenamefont{Yamamoto, Koashi,
  \"{O}zdemir, and Imoto}}]{tyam03}
\bibinfo{author}{\bibfnamefont{T.}~\bibnamefont{Yamamoto}},
  \bibinfo{author}{\bibfnamefont{M.}~\bibnamefont{Koashi}},
  \bibinfo{author}{\bibfnamefont{S.~K.} \bibnamefont{\"{O}zdemir}},
  \bibnamefont{and} \bibinfo{author}{\bibfnamefont{N.}~\bibnamefont{Imoto}},
  \bibinfo{journal}{Nature} \textbf{\bibinfo{volume}{421}},
  \bibinfo{pages}{343} (\bibinfo{year}{2003}).

\bibitem[{woo()}]{wooters98coffman00}
\emph{\bibinfo{title}{{\rm W.~K.~Wooters, Phys.~Rev.~Lett. {\bf 80}, 2245
  (1998); V.~Coffman and J.~Kundu and W.~K.~Wooters, Phys.~Rev.~A {\bf 61},
  052306 (2000).}}}

\bibitem[{\citenamefont{James et~al.}(2001)\citenamefont{James, Kwiat, Munro,
  and White}}]{james01}
\bibinfo{author}{\bibfnamefont{D.~F.~V.} \bibnamefont{James}},
  \bibinfo{author}{\bibfnamefont{P.~G.} \bibnamefont{Kwiat}},
  \bibinfo{author}{\bibfnamefont{W.~J.} \bibnamefont{Munro}}, \bibnamefont{and}
  \bibinfo{author}{\bibfnamefont{A.~G.} \bibnamefont{White}},
  \bibinfo{journal}{Phys.~Rev.~A} \textbf{\bibinfo{volume}{64}},
  \bibinfo{pages}{052312} (\bibinfo{year}{2001}).

\bibitem[{foo({\natexlab{c}})}]{footnotedech}
\emph{\bibinfo{title}{{\rm The optical path length difference of the decoherers
  is not generally exactly 140$\lambda$, causing an extra phase on the off
  diagonal elements. The phase is set to zero by slightly tipping one of the
  decoherers about its vertical axis}}}.

\bibitem[{ber()}]{berg}
\emph{\bibinfo{title}{{\rm A.~J.~Berglund, Dartmouth College B.A. Thesis, also
  quant-ph/0010001 (2000); N.~Peters{ \it et~al.}, J. Quant. Inf. Comp. {\bf
  3}, 503 (2003).}}}

\bibitem[{bar()}]{barbieri03}
\emph{\bibinfo{title}{{\rm As recently demonstrated [M.~Barbieri,
  F.~De~Martini, G.~Di~Nepi, and P.~Mataloni, quant-ph/0303018 (2003);
  G.~Di~Nepi, F.~De~Martini, M.~Barbieri, and P.~Mataloni, quant-ph/0307204
  (2003)], one could instead use the spatial degree of freedom to induce
  decoherence; however, the states are then not suitable, e.g., for use in
  fiber optic systems, or where interference methods are needed.}}}

\bibitem[{\citenamefont{Jozsa}(1994)}]{jozsa94}
\bibinfo{author}{\bibfnamefont{R.}~\bibnamefont{Jozsa}},
  \bibinfo{journal}{J.~Mod.~Optics} \textbf{\bibinfo{volume}{41}},
  \bibinfo{pages}{2315} (\bibinfo{year}{1994}).

\bibitem[{foo({\natexlab{d}})}]{footnotesensitivity}
\emph{\bibinfo{title}{{\rm The leading order normalized behaviors for the
  measures about a target value $\rho_{MEMS i}(r_0)\equiv \rho_{i}(r_0)$ by an
  amount $\delta r\equiv r-r_0$ are
  $S_{L}(\rho_{i}(r))/S_{L}(\rho_{i}(r_0))\approx 1-A_{i} \delta r$,
  $T(\rho_{i}(r))/T(\rho_{i}(r_0))\approx 1+\frac{2}{r_0} \delta r$, and
  $F(\rho_{i}(r_0),\rho_{i}(r))\approx 1+B_{i}(\delta r)^2$, where the
  subscript i denotes the class of MEMS, and the constants are given by
  $A_I=\frac{2 r_0-1}{r_0 (1-r_0)}$, $A_{II}=\frac{2 r_0}{\frac{4}{3}-r_0^2}$,
  $B_I=\frac{-1}{4 r_0 (1-r_0)}$, and $B_{II}=\frac{3}{2(9 r_0^2-4)}$.}}}

\bibitem[{twi()}]{twirl}
\emph{\bibinfo{title}{{\rm In ``twirling,'' a random SU(2) rotation is
  independently performed on each photon pair.}}}

\bibitem[{foo({\natexlab{e}})}]{footnoteeff}
\emph{\bibinfo{title}{{\rm Because it is presently very difficult to produce
  simultaneous indistinguishable {\it pairs} of photons, the filtration
  technique is {\it much} more efficient, e.g., where typically 20\% of our
  incident ensemble of pairs survived, less than 0.005\% would survive
  (estimated from the 2-fold and 4-fold coincidence data reported
  in~\cite{pan}) in the interference schemes, which require 4 photons.}}}

\end{thebibliography}
\vspace{0cm}
\end{document}